\documentclass[aps,pre,floatfix,amsmath,amssymb,superscriptaddress,twocolumn,footinbib]{revtex4-1}
\usepackage{epsfig}
\usepackage{natbib}
\usepackage{color}

\begin{document}

\title{Overdamped dynamics of particles with repulsive power-law interactions}

\author{Andr\'e A. Moreira}
\affiliation{Departamento de F\'isica, Universidade Federal do
 Cear\'a, 60451-970 Fortaleza, Brazil}
\affiliation{National Institute of Science
and Technology of Complex Systems, Rua Xavier Sigaud 150, 22290-180
Rio de Janeiro, RJ, Brazil}
\author{C\'esar M. Vieira}
\affiliation{Departamento de F\'isica, Universidade Federal do
 Cear\'a, 60451-970 Fortaleza, Brazil}
\affiliation{Instituto Federal de Educa\c{c}\~ao, Ci\^encia e Tecnologia
  do Cear\'a, 62580-000 Acara\'u, Brazil}
\author{Humberto A. Carmona}
\affiliation{Departamento de F\'isica, Universidade Federal do
 Cear\'a, 60451-970 Fortaleza, Brazil}
\author{Jos\'e S. Andrade, Jr.}
\affiliation{Departamento de F\'isica, Universidade Federal do
  Cear\'a, 60451-970 Fortaleza, Brazil}
\affiliation{National Institute of Science
and Technology of Complex Systems, Rua Xavier Sigaud 150, 22290-180
Rio de Janeiro, RJ, Brazil}
\author{Constantino Tsallis}
\affiliation{Centro Brasileiro de Pesquisas F\'isicas, Rua Xavier Sigaud 150,
22290-180 Rio de Janeiro, RJ, Brazil}
\affiliation{National Institute of Science
and Technology of Complex Systems, Rua Xavier Sigaud 150, 22290-180
Rio de Janeiro, RJ, Brazil}
\affiliation{Santa Fe Institute, 1399 Hyde Park Road,
  New Mexico 87501, USA}
\affiliation{Complexity Science Hub Vienna, Josefstadter Strasse 39, 1080 Vienna, Austria}

\begin{abstract}
  We investigate the dynamics of overdamped $D$-dimensional systems
  of particles repulsively interacting through
  short-ranged power-law potentials, $V(r)\sim r^{-\lambda}\;(\lambda/D>1)$. 
  We show that such systems obey a non-linear diffusion
  equation, and that their stationary state extremizes a $q$-generalized nonadditive
  entropy. Here we focus on the dynamical evolution of these systems. 
  Our first-principle $D=1,2$ many-body numerical simulations (based on Newton's law) confirm the
  predictions obtained from the time-dependent solution of the
  non-linear diffusion equation, and show that the one-particle
  space-distribution $P(x,t)$ appears to follow a compact-support $q$-Gaussian form, with $q=1-\lambda/D$. 
  We also calculate the velocity distributions $P(v_x,t)$ and, interestingly enough, they follow the same $q$-Gaussian form (apparently precisely for $D=1$, and nearly so for $D=2$). The satisfactory match between the continuum
  description and the molecular dynamics simulations in a more general,
  time-dependent, framework neatly confirms the idea that the present
  dissipative systems indeed represent suitable
  applications of the $q$-generalized thermostatistical theory.
\end{abstract}\maketitle

\section{Introduction}

\begin{figure}[t]
\centerline{\psfig{file=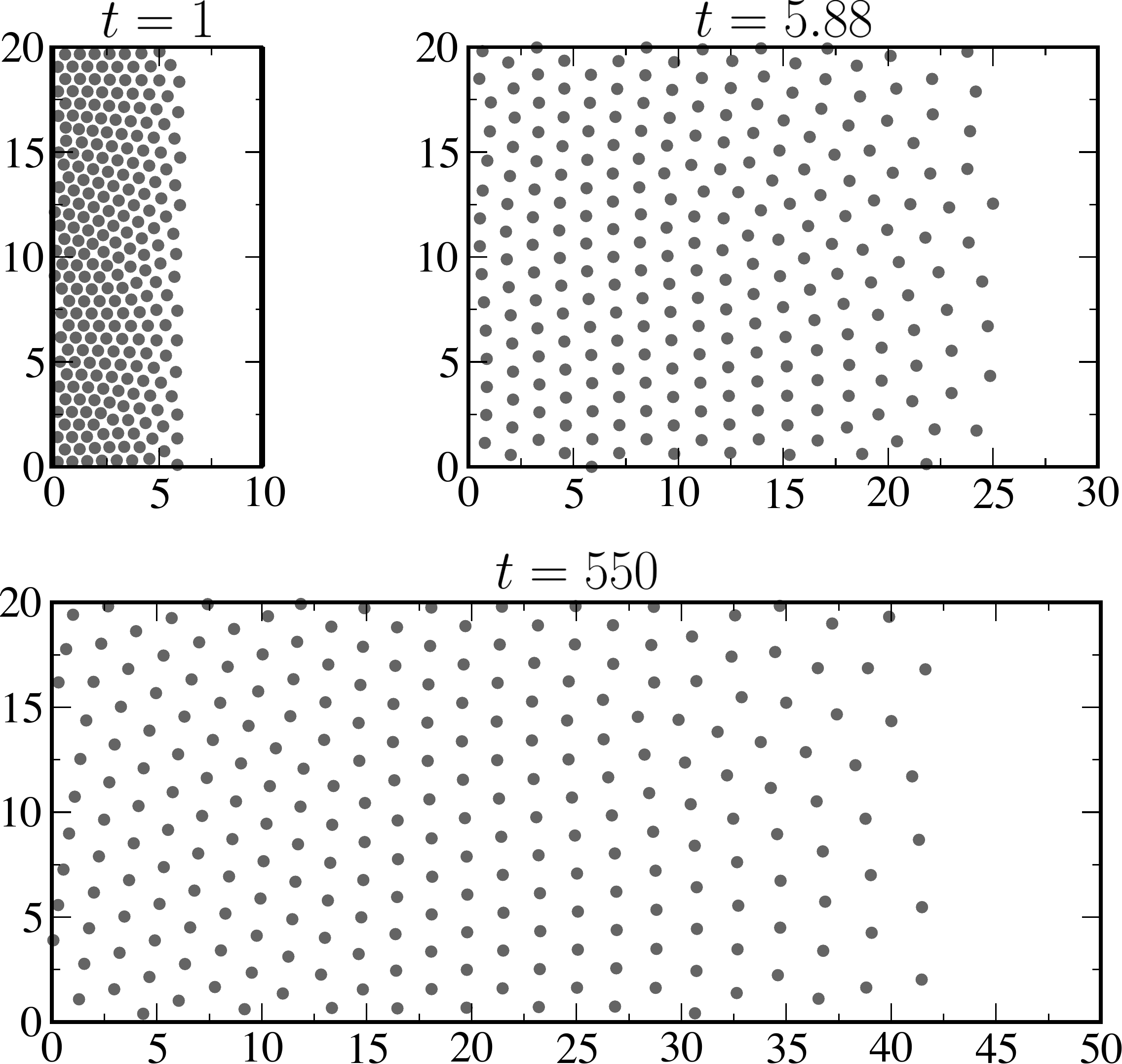, width=1\linewidth}}
\caption{Snapshots of the configuration of the system in three
  different moments of the dynamics. Due to the symmetry we show
  only the $x>0$ half of the system. The particles start
  in a very narrow region and invade the system as time goes, eventually
  reaching an equilibrium position. This system
  comprises $N=500$ particles, interacting through a potential
  $V_{ij}=\epsilon(r_{ij}/\sigma)^{-\lambda}$, with $\lambda=4$,
  in a cell of lateral length $L_y=20\sigma$, and
  confined by an external force $-kx$  with $k=1\times10^{-6}\epsilon/\sigma^2$.
  }
\label{fig0}
\end{figure}

Dissipative systems of repulsive particles are representative of
many physical phenomena in nature, including for instance, type-II
superconductors~\cite{gennes, ref8, *ref9_1, *ref9_2, *ref9_3,
  ref10_1, *ref10_2, *ref11_1, *ref11_2, ref12, *ref13_1, *ref13_2,
  zapperi2001flux, george, petrucio, *prb_moreira}, complex
plasmas~\cite{shukla,hiroo}, and colloidal
systems~\cite{diego1,*diego2,munarin,peeters}. In the overdamped limit, the
equations of motion for such systems take the form of a first order
differential equation, where the velocity of the particles is
proportional to the force over them,
$\mathbf{v}_i=\mathbf{F}_{i}/\gamma$. A recent work~\cite{cesar} has
shown that for a wide variety of possible repulsive potentials the
local density $\rho(\mathbf{r},t)$ of these overdamped repulsive
particles should follow a nonlinear diffusion equation in the form
\begin{equation}\label{eq:model}
  \gamma\frac{\partial \rho}{\partial t}=\boldsymbol{\nabla}\cdot\left\lbrace\rho\left[\boldsymbol{\nabla}U_{ext}+a(\rho)\boldsymbol{\nabla}\rho\right]\right\rbrace,
\end{equation}
where $U_{ext}(\mathbf{r})$ is an applied external
potential.

The function $a(\rho)$ can be obtained from the potential energy
$U_1$ of a particle in the homogeneous state of density $\rho$~\cite{cesar},
\begin{equation}\label{arho}
  a(\rho)=2\frac{dU_1}{d\rho}+\rho\frac{d^2U_1}{d\rho^2}.
\end{equation}
Determining $U_1$ depends on the knowledge of the interaction
potential and on the microscopic structure in which the particles rest
in the homogeneous state~\cite{cesar}.  As mentioned, the
applicability of this approach is restricted to systems of overdamped
particles interacting through a short-ranged repulsive
potential. More precisely, for large distances $r$, the
interaction potential $V(r)$ should decay faster than $r^{-D}$, where
$D$ is the dimensionality of the system. The form of $a(\rho)$ is also
influenced by the way the potential diverges at the origin. If the
potential diverges at the origin slower than $r^{-D}$, $a(\rho)$
should converge to a finite value for densities $\rho$ that are
sufficiently large. Conversely, for potentials that diverge faster
than $r^{-D}$ at the origin, the energy per particle $U_1$ grows
faster than linearly with the density $\rho$, and $a(\rho)$ never
converges to a fixed value, regardless of the density
$\rho$~\cite{cesar}. In the case where the interaction potential is a
power law, $V_{ij}=\epsilon(r_{ij}/\sigma)^{-\lambda}$, with
$\lambda>D$, $\sigma >0$ and $\epsilon >0$, Eq.~(\ref{eq:model}) can be written as
\begin{equation}\label{plmod}
  \gamma\frac{\partial \rho}{\partial t}=\boldsymbol{\nabla}\cdot\left[\rho\left(\boldsymbol{\nabla}U_{ext}+C_\lambda\rho^{\frac{\lambda}{D}-1}\boldsymbol{\nabla}\rho\right)\right],
\end{equation}
where the constant $C_\lambda$ can be computed from the structure of the homogeneous state~\cite{cesar}.
More recently, for this family of repulsive potentials, a
consistent thermodynamic framework was developed, and thermodynamic
potentials, Maxwell relations, and response functions could be
obtained~\cite{souza}.

Here, we analytically obtain time-dependent solutions for
Eq.~(\ref{plmod}), $\rho(x,t)$, for the case of a parabolic confining
potential, $U_{ext}=-kx^2/2$ ($k>0$), showing that they possess the form of
$q$-Gaussian distributions~\cite{tsallis, *tsallis1, *tsallis2, adib,
  zipf, murilo}, as well as the dynamics of type-II superconducting
vortices~\cite{ribeiro, nobre2}.  We compare these solutions with the
results obtained from computer simulations, where the equations of
motion are solved numerically, and find a good agreement. Moreover,
our results show that, for the class of solutions we obtain, the
velocities of the particles should be proportional to their position,
that is, $v_i\sim x_i$, as also observed in Ref.~\cite{ribeiro} for
the London potential. This suggests that the velocity distribution
should follow closely the same expected form for the density profile,
namely, a $q$-Gaussian. As we show, however, due to the readjustment
of the local spatial structure of the system as the surrounding density
changes, it generates, for $D>1$, an extra noise that leads, as compared to $q$-Gaussians, to small deviations in
the shape of the velocity distributions at the highest velocities. 

\section{Model Solution}

\begin{figure}[t]
\centerline{\psfig{file=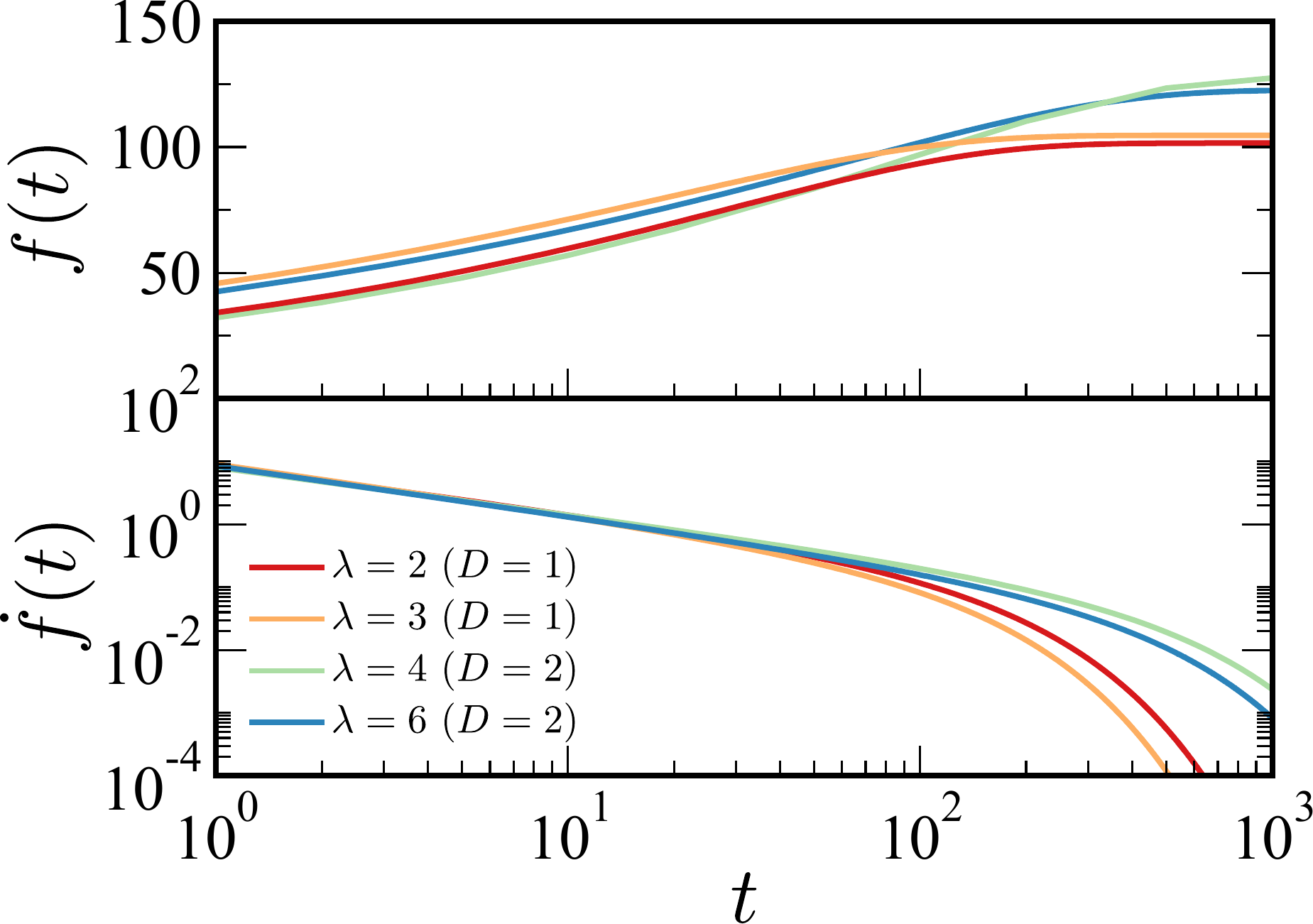, width=1\linewidth}}
\caption{ Curves of $f(t)$ for $D=1$ ($\lambda=2$ and 3) and $D=2$
  ($\lambda=4$ and 6) obtained from Eq.~(\ref{eq:solf}) and its
  derivative $\dot{f}(t)$. In our numerical simulations we start
  with all particles confined in a narrow stripe, leading us to use
  $t_0=0$ in Eq.~(\ref{eq:solf}). The case $D=1$ corresponds to
  $N=3600$ particles with confining potential strength
  $k=3.2\times10^{-3}\epsilon/\sigma^2$.  The case $D=2$ corresponds
  to $N=4000$ particles, with confining potential strength
  $k=1\times10^{-3}\epsilon/\sigma^2$, in a cell with transverse size
  $L_y=20\sigma$.  }
\label{fig1}
\end{figure}

In the present work, the systems that we model consist of $N$
particles interacting through the above mentioned potential
$V_{ij}=\epsilon(r_{ij}/\sigma)^{-\lambda}$ and
confined in the $x$-direction by the external potential $U_{ext}$. For the case of two dimensions, in the
$y$-direction the simulation cell has a finite length $L_y$ and
periodic boundary conditions are imposed. See Fig.~\ref{fig0} for a view
of a 2D system in different moments of the dynamics. In this case, it is
reasonable to expect the density to be independent on $y$, so that
Eq.~(\ref{plmod}) can be written as
\begin{equation}\label{pl1d}
  \gamma\frac{\partial \rho}{\partial t}=\frac{\partial}{\partial x}\left[\rho\left(kx+C_\lambda\rho^{\frac{\lambda}{D}-1}\frac{\partial\rho}{\partial x}\right)\right].
\end{equation} One may find solutions of Eq.~(\ref{pl1d}) by making use of the
similarity
hypothesis~\cite{bryksin}, \begin{equation}\label{hhh}\rho(x,t)=\frac{g(z)}{f(t)},\end{equation}
with $z=x/f(t)$. Using this in Eq.~(\ref{pl1d}), we obtain
\begin{equation}\label{aaa}
\frac{f^{1+\frac{\lambda}{D}}}{C_\lambda}\left(\frac{df}{dt} + kf\right)=-\frac{\frac{d}{dz}\left(g^{\frac{\lambda}{D}}\frac{dg}{dz}\right)}{\frac{d}{dz}(gz)}.
\end{equation}
The left side of Eq.~(\ref{aaa}) depends on $t$, while the right side
depends on $z$. The only possible solution is that both sides are equal to some
constant, $\nu$. From that, the left side becomes
\begin{equation}\label{eq:left}
\frac{df}{dt}=\frac{\nu C_\lambda}{f^{\frac{\lambda}{D}+1}}-kf
\end{equation}
while the right side can be written as
\begin{equation}\label{eq:right}
\frac{d}{dz}\left(g^{\frac{\lambda}{D}}\frac{dg}{dz}+\nu gz\right)=0.
\end{equation}
To solve  Eq.~(\ref{eq:right}) we consider the boundary condition $g(1)=0$, leading to
\begin{equation}\label{eq:solg}
g=\left[\frac{\lambda\nu}{2D}(1-z^2)\right]^{\frac{D}{\lambda}},
\end{equation}
%
%
that is, the shape of the density profile is a $q$-Gaussian with $q=1-\lambda/D$ at
any instant of time.  The normalization condition,
$\int\rho(x,t)dx=n$, leads to $\int{gdz} =n$, where $n=N$ for $D=1$
and $n=N/L_y$ for $D=2$,
remembering that
$L_y$ gives the thickness of the simulation cell and $N$ the number of
particles. This allows us to determine the value of $\nu$, namely
\begin{equation}\label{eq:alp}
\nu=\frac{2D}{\lambda}\left[n\frac{\Gamma(\tfrac{3}{2}+\tfrac{D}{\lambda})}{\Gamma(1+\tfrac{D}{\lambda})\sqrt{\pi}}\right]^{\frac{\lambda}{D}}.
\end{equation}
It is also visible from this solution that $f(t)$ is the point where
$\rho(f(t),t)=0$, that is, $f(t)$ is the edge of the
distribution. Solving Eq.~(\ref{eq:left}) we obtain
\begin{equation}\label{eq:solf}
f(t)=\left\lbrace\frac{\nu C_\lambda}{k}\left[ 1 - e^{-k\left(\frac{\lambda}{D}+2\right)(t-t_0)}\right] \right\rbrace^{\frac{D}{\lambda+2D}},
\end{equation}
where $t_0$ is a free parameter that depends on the initial
condition. Since in our numerical simulations we start with all
particles confined in a narrow stripe, $t_0=0$ should fit well 
our numerical results.  In Fig.~\ref{fig1} we show the curves of $f(t)$ and
$\dot{f}(t)\equiv df/dt$ for each instant of time $t$ considering
$\lambda=2$ and $\lambda=3$ in one dimension, and $\lambda=4$ and
$\lambda=6$ in two dimensions. These curves were obtained from
Eq.~(\ref{eq:solf}). As we show in what follows, these predictions
closely agree with the results from the numerical simulations.

\section{Numerical Simulations}

\begin{figure}[t]
\centerline{\psfig{file=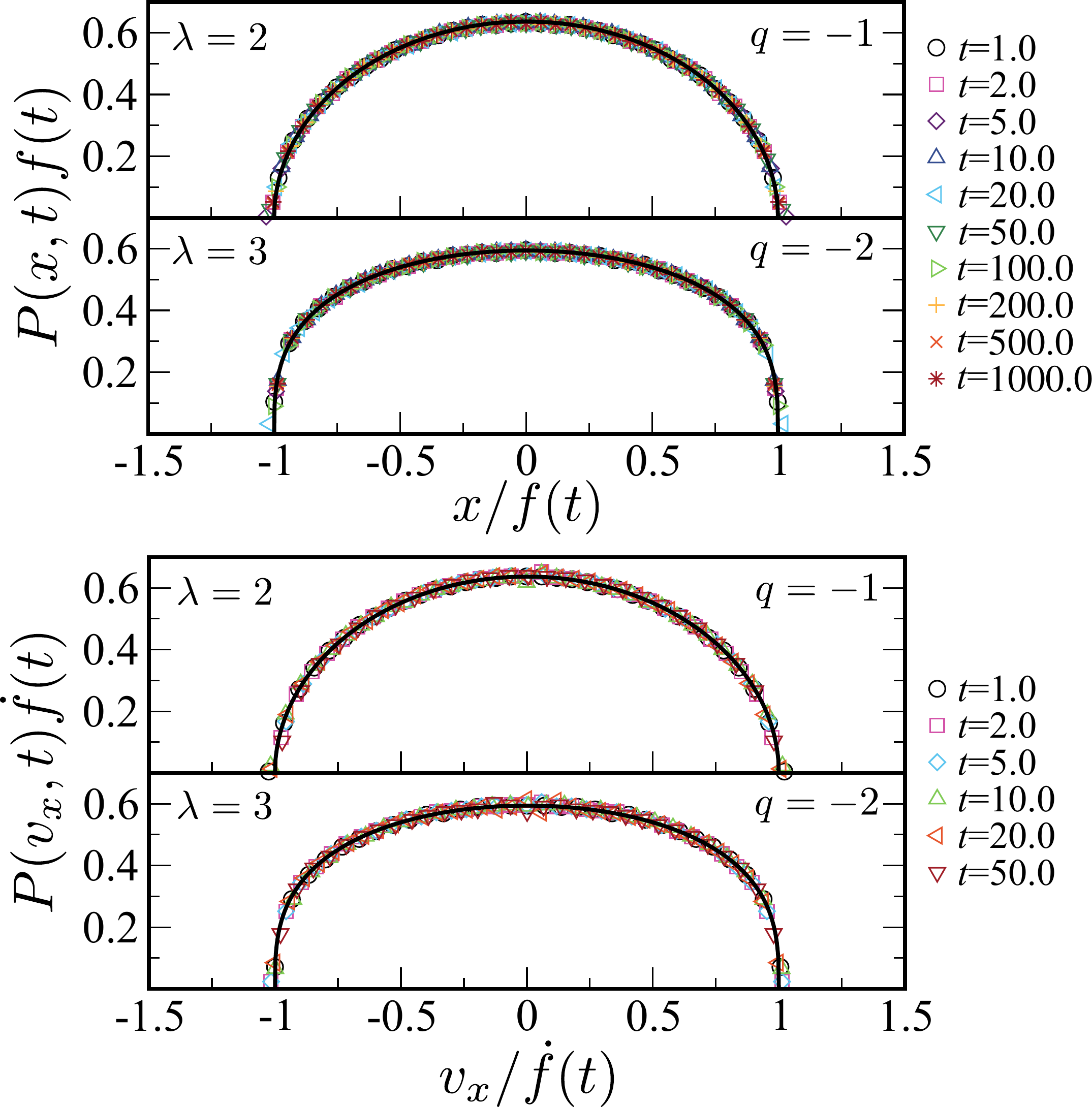, width=1.0\linewidth}}
\caption{Distributions of scaled positions and velocities at different
  moments of the dynamics for the one dimensional case ($D=1$). These
  results concern $N=3600$ particles interacting through the power law
  potential, $V=\epsilon(r/\sigma)^{-\lambda}$, for
  $\lambda=2$~($q=-1$) and $\lambda=3$~($q=-2$).  The black curves are
  $q$-Gaussians. Here we used the strength of the confining potential
  $k=3.2\times10^{-3}\epsilon/\sigma^2$.
}
\label{fig2}
\end{figure}

Figure~\ref{fig2} shows the density profile at different moments of
the dynamics for one dimensional systems. To obtain these curves,
we performed the Kernel Density Estimation~\cite{parzenkde} for the
position of the particles scaled by the length $f(t)$ obtained from
Eq.~(\ref{eq:solf}). 
The results from simulation are in perfect agreement with the
predicted form given by Eq.~(\ref{eq:solg}), showing that, in fact,
Eq.~(\ref{eq:solf}) yields the correct position $f(t)$ at the edge of
the density profile.

Next we proceed to investigate the velocity of particles during the
dynamics. To obtain a solution for our non-linear diffusion,
Eq.~(\ref{pl1d}), and considering the similarity hypothesis,
Eq.~(\ref{hhh}), from Eq.~(\ref{eq:solg}) it is visible that
$g(1)=0$, that is, $f(t)$ is the position where the density profile
vanishes.  It is reasonable to assume that the average velocities of
the particles at a given position and time, $\bar{v}(x,t)$, also obey
Eq.~(\ref{hhh}),
\begin{equation}
\bar{v}(x,t)=\dot{f}(t)b(z),~\text{with}~\dot{f}(t)=df/dt.
\end{equation}
The average velocity of the particles at the edge of the density
profile is given by $\bar{v}(f(t),t)=\dot{f}(t)$.  By inserting the
similarity hypothesis into the continuity equation, we obtain
\begin{equation}
-\frac{\dot{f}}{f^2}\frac{d}{dz}\left(gz
  \right)=-\frac{\dot{f}}{f^2}\frac{d}{dz}\left(gb\right),
\end{equation}
leading to the condition that $\bar{v}(x,t)=(\dot{f}/f)x$, that is,
the average velocity is linear with position.

In one dimension there is not more than one particle at each given position
$x$, therefore if $\bar{v}(x,t)$ is linear with $x$, the velocity of
each particle should be linear with $x$, leading to the
conclusion that they are distributed in the same form, namely, a
$q$-Gaussian. This prediction is consistent with~\cite{plastino18},
and  confirmed by the
results of Fig.~\ref{fig2}. In larger dimensionalities
there are several particles in the stripe around a given distance
$x$. In this case $\bar{v}(x,t)$ may still be linear with $x$, but
the velocities of each particle may fluctuate around this average.

\begin{figure}[t]
\centerline{\psfig{file=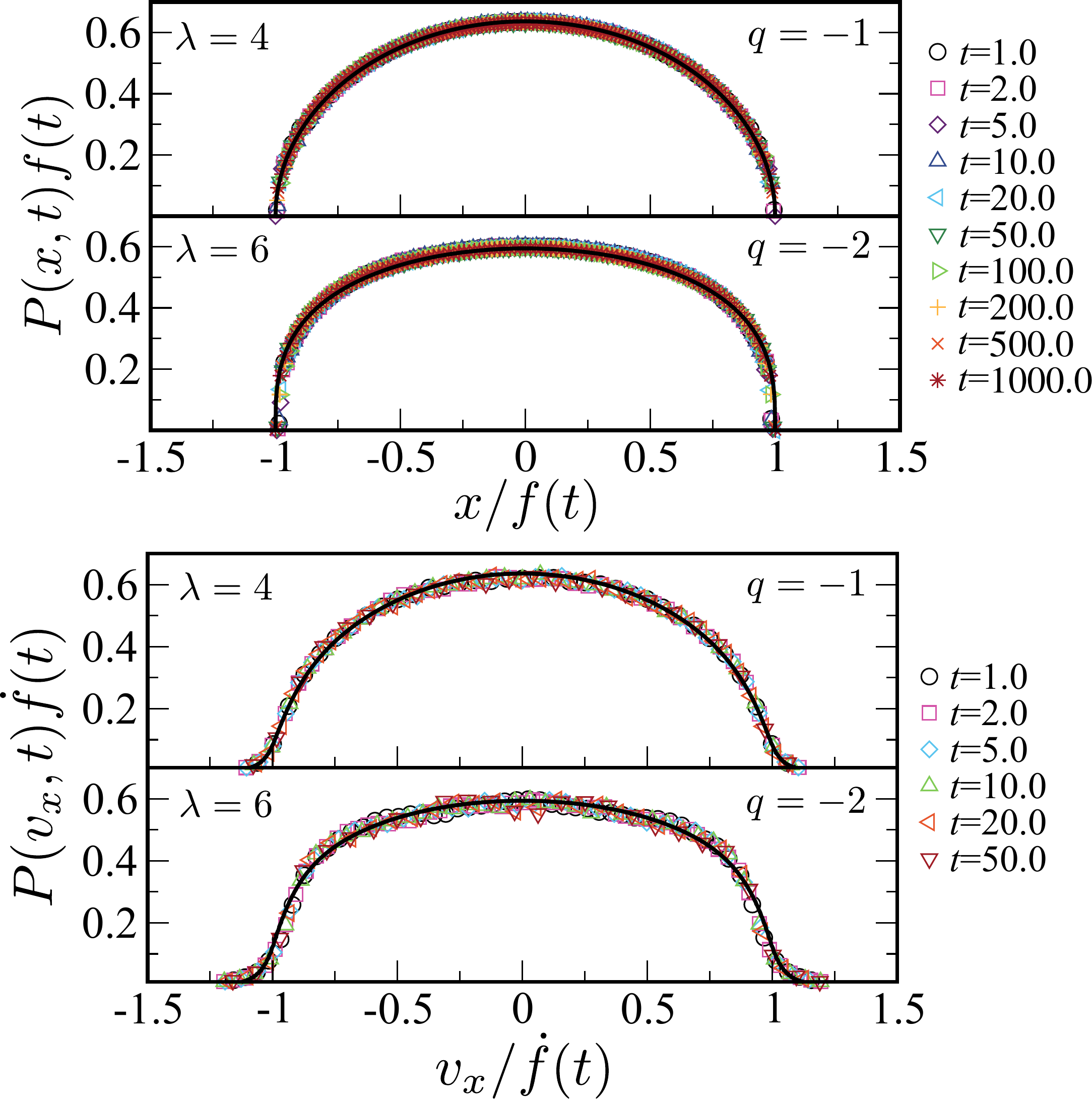, width=1.0\linewidth}}
\caption{ Distributions of scaled positions and velocities at
  different moments of the dynamics for the case of two dimensions
  ($D=2$). These results concern particles interacting through the
  power law potential, $V=\epsilon(r/\sigma)^{-\lambda}$, for
  $\lambda=4$~($q=-1$) and $\lambda=6$~($q=-2$).  One can compare the
  density profiles with $q$-Gaussians (black curves). In the case of
  the velocity distributions the black curves show convolutions
  between $q$-Gaussians and Laplacian distributions, as given by
  Eq.~(\ref{eq:pvx}), with the parameter $h$ given by 0.045 and 0.051
  for $\lambda=4$ and 6, respectively. In these simulations we used
  $N=4000$ particles in a cell of transverse size $L_y=20\sigma$ with
  periodic boundary conditions. In the longitudinal direction ($x$
  axis) we imposed a confining force $-kx$ with
  $k=1\times10^{-3}\epsilon/\sigma^2$.
%
}
\label{fig3}
\end{figure}

\begin{figure}[t]
\centerline{\psfig{file=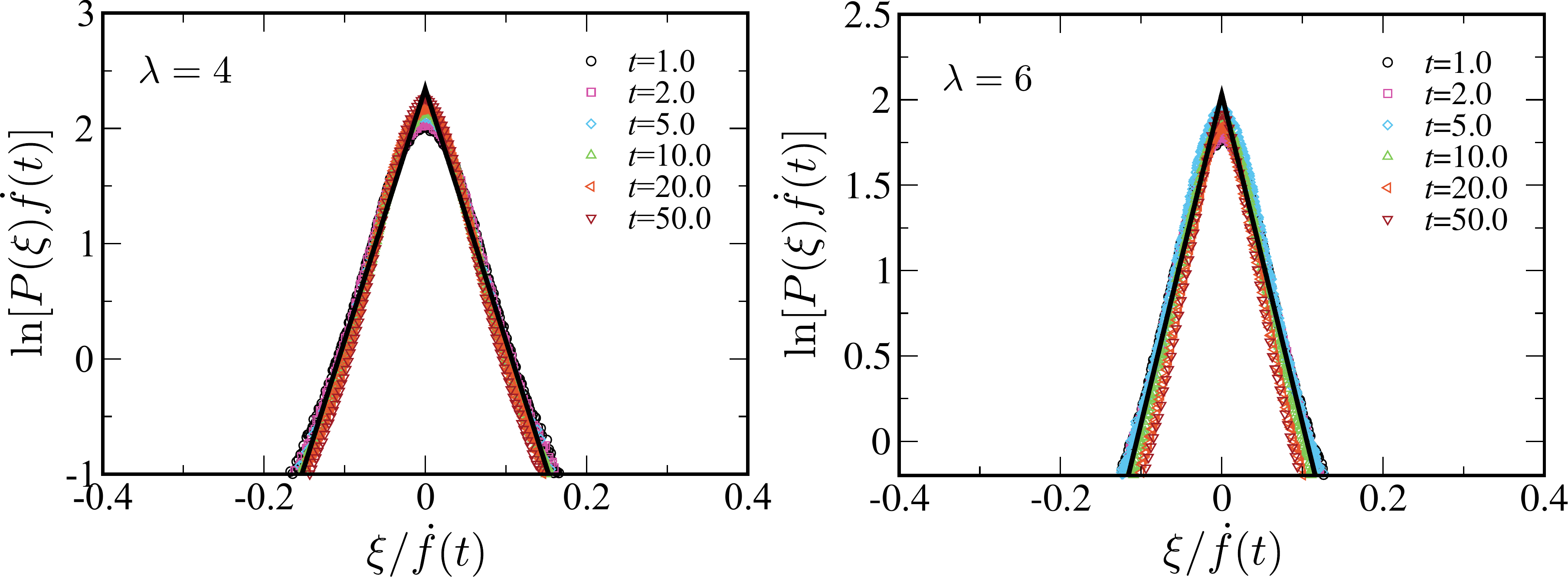, width=1.0\linewidth}}
\caption{Distributions of the displacement of the average
  velocity, $\xi_i=(v_x)_i/\dot{f} - x_i/f$ for some values of times
  $t$ considering $\lambda=4$ and $\lambda=6$ for $D=2$. The black curves represent
  Laplacian distributions, $P(\xi)\sim \exp(-|\xi|/h)$, where $h$ is
  an adjustment parameter, given by 0.045 and 0.051 for $\lambda=4$
  and 6, respectively.  }
\label{fig4}
\end{figure}

To test this hypothesis we performed simulations in two dimensions.
Figure \ref{fig3} presents the density profiles and velocity
distributions obtained from simulations. As before, positions where
scaled by $f(t)$, while velocities where scaled by $\dot{f}(t)$.  As
exhibited, the density profiles follow closely the $q$-Gaussian form. 
The velocity distributions also display an invariant shape for
all instants. However, this shape deviates slightly from the expected
$q$-Gaussian form at the largest velocities. Note that to reduce fluctuations in our results we
performed averages over 800 sample simulations.
To investigate this small difference, we analyze the distribution of the
quantity $\xi_i\equiv (v_x)_i/\dot{f} - x_i/f$ among the particles of
all samples, which measures how far $(v_x)_i$ is from the expected
average $\bar{v}(x_i,t)=(\dot{f}/f)x_i$ for a particle at position
$x_i$ at time $t$. Figure~\ref{fig4} shows the distributions
$P(\xi)$ for different values of $t$. One can see that $P(\xi)$ can be
described approximately by a Laplacian distribution,
\begin{equation}\label{eq:laplace}P(\xi)\sim\exp(-|\xi|/h),\end{equation} where the parameter $h$ depends on the
particular value of $\lambda$. Most likely, these small deviations in the
expected value of the velocity are due to the rearrangement of 
the local spatial structure as the density changes.
As we know from the results of Fig.~\ref{fig4}, the distribution of
positions can be well described as a $q$-Gaussian, $P(x/f)=G_q(x/[(q-1)f])$. Since $f(t)$ and $\dot{f}(t)$ depend only on time, and
$(v_x)_i/\dot{f}=x_i/f + \xi_i$, then the $v_x$-distribution can be
found by the convolution~\cite{hogg} between a $q$-Gaussian and a
Laplacian distribution, 
%
\begin{equation}\label{eq:pvx}
  P\left(\frac{v_x}{\dot{f}}\right)=\frac{1}{2h} \int_{-\infty}^\infty d\xi~ G_q\left(\frac{v_x/\dot{f}-\xi}{q-1}\right)e^{-\frac{|\xi|}{h}}.
\end{equation}
The black curves shown in the Fig.~\ref{fig3} have been obtained from
Eq.~(\ref{eq:pvx}), where the values of $h$ were obtained from the best fits to the molecular simulation data of the Laplace distribution, Eq.~(\ref{eq:laplace}), as shown in Fig.~\ref{fig4}.

\section{Discussion}

We have studied a system of particles interacting through power-law
repulsive potentials, and under overdamped motion. In a previous work,
through a coarse-graining approximation, this model was related to a
nonlinear diffusion equation, whose stationary-state solutions have
been shown to be compatible with results obtained from
molecular-dynamics simulation~\cite{cesar}. Here, we investigate the whole time evolution. Using a similarity
hypothesis, we showed that our nonlinear diffusion equation predicts
that, for all times, the probability distribution for the positions,
$P(x,t)$, is a $q$-Gaussian with the value of $q$ depending on both
the repulsive potential as well as on dimensionality of the system, 
$q=1-\lambda/D$. We present quite satisfactory results from molecular
dynamics simulations to give support to the analytic predictions.
Moreover, we have also presented results for the $x$-component
velocity probability distribution, $P(v_x ,t)$, showing that it is
given by a $q$-Gaussian distribution that, in larger dimensionalities,
will be perturbed by a small extra noise well approximated by a Laplacian
distribution. We conjecture that these perturbations are due to the
rearrangement of the local spatial structure as the density
changes.  To summarize, we have presented broad evidence that a system
of overdamped repulsive particles interacting through a short-range
power-law potential constitutes an important physical application for
nonextensive statistical mechanics. Both stationary
states and time-dependent properties of the systems are fully
compatible with the theory.

\begin{acknowledgments}
The authors thank the Brazilian agencies CNPq, CAPES, FUNCAP, and the
National Institute of Science and Technology for Complex Systems
(INCT-SC) in Brazil for financial support.
\end{acknowledgments}

\bibliographystyle{apsrev4-1.bst}
\bibliography{biblio}

\end{document}